\documentclass[aip,jcp,reprint,eqsecnum,superscriptaddress,twocolumn,longbibliography,floatfix]{revtex4-2}

\usepackage{empheq}
\usepackage{graphicx}
\usepackage{dcolumn}
\usepackage{bm}

\usepackage[utf8]{inputenc}
\usepackage[T1]{fontenc}
\usepackage{mathptmx}
\usepackage{etoolbox}
\usepackage{xcolor}
\usepackage{mathpazo}
\usepackage[unicode=true,
 bookmarks=true,bookmarksnumbered=false,bookmarksopen=false,
 breaklinks=false,pdfborder={0 0 0},pdfborderstyle={},backref=false,colorlinks=true]
 {hyperref}
\hypersetup{pdfborderstyle={},linkcolor=blue,citecolor=blue}
\usepackage{breakurl}

\usepackage{natbib}
\usepackage{amssymb,amsmath}

\usepackage{float}

\newcommand{\rr}{\mathbf{r}}
\newcommand{\RR}{\mathbf{R}}

\newcommand{\td}{\theta_{12}}
\newcommand{\xd}{x_{12}}
\newcommand{\rd}{r_{12}}
\newcommand{\Rd}{R_{12}}
\newcommand{\PP}{\mathcal{P}}

\newcommand{\smin}{s_{\min}}

\makeatletter
\def\@email#1#2{%
 \endgroup
 \patchcmd{\titleblock@produce}
  {\frontmatter@RRAPformat}
  {\frontmatter@RRAPformat{\produce@RRAP{*#1\href{mailto:#2}{#2}}}\frontmatter@RRAPformat}
  {}{}
}%
\makeatother
\begin{document}

\title[]{On the correlation lengths of confined spheres in a cylindrical pore}
\author{Ana M. Montero}
\affiliation{Departamento de F\'isica, Universidad de Extremadura, E-06006 Badajoz, Spain}
\email{anamontero@unex.es}

\date{\today}

\begin{abstract}
We investigate the structural correlations of hard spheres confined within a narrow cylindrical pore in the quasi-one-dimensional regime, where interactions are restricted to nearest neighbors. Using a Laplace-space formulation of the radial distribution function (RDF), we determine the correlation lengths and oscillation frequencies associated with its long-distance decay. In addition to the global RDF, we analyze transverse-resolved RDFs that account for the positions of particle pairs across the pore cross section. While these observables are associated with the same underlying pole spectrum, their residues depend on the transverse configuration and can vanish due to symmetry. As a result, different particle-pair configurations may be governed by different leading poles and display different correlation lengths and oscillation frequencies. In particular, the global RDF does not always reflect the longest-ranged correlations found in transverse-resolved observables. We examine how this behavior depends on density and confinement. In the strong-confinement limit, the system approaches the Tonks-gas behavior at finite pressure, and the differences between the RDFs disappear.
\end{abstract}

\maketitle

\section{\label{sec:introduction}Introduction}

Understanding the correlation length in systems of interacting particles is central to statistical mechanics because it quantifies how microscopic fluctuations propagate across space and influence macroscopic behavior,~\cite{ELHH94,DE00,SD05,SHER19} connecting microscopic interactions to collective behavior such as criticality or crystallization.~\cite{PZVSPC09,WSAE18} In general, the correlation length characterizes the distance over which particle positions or fluctuations remain statistically correlated and its analysis for a wide variety of systems has been a central part of the study of liquids for years.~\cite{K39,TS77,SD05,HS10}
In the context of hard spheres, the correlation length also reflects how excluded-volume interactions alone can generate nontrivial structural organization.~\cite{PT72,GDER04,PBYSH20,PYSHB21,SYH20,PZVSPC09}

Although one-dimensional systems with short-range interactions do not exhibit true thermodynamic phase transitions,~\cite{vH50, CS04} the study of the correlation length in these systems remains highly informative, since it encodes how rapidly spatial correlations decay and also provides a precise benchmark for theoretical approaches.~\cite{KT68,FW69,SFG08,FGMS10,MS19,BM25} Because many one-dimensional models are exactly solvable, they offer a controlled setting to test analytical techniques and validate numerical algorithms.

Quasi-one-dimensional (Q1D) systems are those in which particles are confined by a narrow geometry that strongly restricts their motion: particles cannot bypass one another and remain in a single-file arrangement along the longitudinal direction, while retaining a limited degree of freedom in the transverse directions.~\cite{GV13,PK92,GM14,RGM16,LZHRL17,MBGM22,JF22,MS24,MS24b,FS24,GV25} The phase behavior of these systems closely resembles that of strictly one-dimensional systems: they do not exhibit true thermodynamic phase transitions, and their correlation length therefore remains finite. Nevertheless, studying the correlation length---particularly how correlations decay in space---remains highly relevant and provides valuable insight into the emergence of structural organization.~\cite{HFC18,MS23b,MS26}

In the strongly confined Q1D regime considered here, particle ordering along the pore axis is preserved and the structural correlations are most naturally described along the longitudinal coordinate. Henceforth, in this context, the term ``radial distribution function'' (RDF) will be understood, for convenience, as referring solely to the longitudinal component.

For spatially confined Q1D systems, and particularly for hard spheres restricted within narrow channels, the correlation length $\xi$ has typically been determined through two main approaches. The first one employs the transfer-matrix method, which yields a correlation length associated with the number of neighbors in between any two correlated particles, rather than as a function of their physical separation distance.~\cite{VBG11,GM15,HFC18,HC21} The second approach relies on numerical simulations of the RDF, from which the correlation length is subsequently obtained by fitting the long-range decay of the RDF to an exponential form.~\cite{HBPT20,HC21} 

On the other hand, in previous studies of Q1D systems, attention has typically been focused on global particle correlations, while partial correlations between particles at different transverse positions within the channel have largely been overlooked. This omission is significant, as it reveals that the structure of correlations in Q1D confinement is not yet fully understood.

Transverse-resolved correlations can provide additional insight into the system, capturing features that are not accessible through the global RDF alone. As such, analyzing these partial correlations is essential for a more complete description of the structural behavior of confined fluids. In fact, some partial correlations can persist over longer distances than those of the global RDF alone, making them especially worth analyzing.

This work focuses on a Q1D system of hard spheres confined within a narrow pore, whose small radius restricts interactions to nearest neighbors. For this system, we use a direct approach based on the analysis of the pole structure of the Laplace transform of the RDF to compute the correlation lengths for both transversely resolved particle pairs and the global RDF, as a function of interparticle longitudinal distance. In addition, we evaluate several other quantities relevant to the study of the asymptotic decay, such as the asymptotic oscillation frequency.

The remainder of this paper is organized as follows: Section~\ref{sec:theory} introduces the confined hard-sphere model and the theoretical framework based on the Laplace-space representation to analyze longitudinal pair correlations and correlation lengths. Section~\ref{sec:numerics} presents a description of numerical methods used for the analysis of the pole and residue structure. The results are presented in Sec.~\ref{sec:results}. Section~\ref{sec:conclusions} summarizes the main conclusions.

\section{\label{sec:theory}Theoretical framework}

\subsection{\label{sec:system}Model description}

We study a system that consists of $N$ identical hard spheres of unit diameter confined inside a narrow three-dimensional cylindrical channel of length $ L \gg 1 $. The diameter of the channel is $ w = 1 + \epsilon $, where the parameter $ \epsilon $ controls the space in the radial direction that is available to the particles' centers. In order to prevent second nearest-neighbor interactions, the excess pore diameter is restricted to $ 0 < \epsilon < \sqrt{3}/2 \simeq 0.866 $. The cylinder’s axis is taken as the $ x $-axis, and its origin can be chosen arbitrarily. A particle’s position vector is then
\begin{equation}
	\RR = x\hat{\mathbf{x}} + \rr,
\end{equation}
where $ -\infty < x < \infty $ is the unconfined longitudinal direction and the transverse component---the position vector along the cross section of the cylinder---is $\rr = y\hat{\mathbf{y}} + z\hat{\mathbf{z}}$. Alternatively, we can also describe this transverse position using polar coordinates $(r,\theta)$, where $ y = r\cos\theta $ and $ z = r\sin\theta $. The accessible region for the sphere centers is $ 0 \leq r \leq \epsilon/2 $ and $ 0 \leq \theta < 2\pi $.

The interaction between any two spheres is defined by the hard-sphere pair potential
\begin{equation}
	\varphi(R_{12}) =
	\begin{cases}
		\infty, & R_{12} < 1, \\
		0, & R_{12} \geq 1,
	\end{cases}
\end{equation}
where $ R_{12} = |\RR_{12}| $ and $ \RR_{12} = \RR_1 - \RR_2 $ is the vector connecting the centers of both spheres.

In general, for two particles located at $\RR_1$ and $\RR_2$, the center-to-center distance is
\begin{equation}
	\Rd = \sqrt{\xd^2 + \rd^2},
\end{equation}
where the longitudinal separation is $\xd = |x_1 - x_2|$ and the transverse distance is
\begin{equation}
	\rd = |\rr_1 - \rr_2| = \sqrt{r_1^2 + r_2^2 - 2r_1r_2\cos\td},
\end{equation}
with $\td = \theta_1 - \theta_2$. When two spheres are in contact ($\Rd = 1$), the longitudinal component of their separation becomes
\begin{equation}\label{eq:contact_distance}
	a_{\rr_1,\rr_2} = \sqrt{1 - \rd^2}.
\end{equation}

In this setting, the number density of the system is
\begin{equation}
	\rho = \frac{N}{L(\pi \epsilon^2 / 4)},
\end{equation}
where only the region available to the sphere centers is considered. However, since the motion along the cylinder is effectively one-dimensional, it is in general more convenient to define a linear density $\lambda = N / L$ that also fully characterizes the density of the system, leading to
\begin{equation}
	\rho = \frac{\lambda}{\pi \epsilon^2 / 4}.
\end{equation}
The minimum contact distance $a_{\mathbf{r}_1,\mathbf{r}_2}=\sqrt{1-\epsilon^2}$ in Eq.~\eqref{eq:contact_distance} occurs for $\theta_{12} = \pi$ and $r_1=r_2=\epsilon/2$, which defines the close-packing linear density as
\begin{equation}
	\lambda_{\text{cp}}(\epsilon) = \frac{1}{\sqrt{1 - \epsilon^2}}.
\end{equation}
This means that the close-packing density of our system ranges from $\lambda_{\text{cp}}(0) = 1$ to $\lambda_{\text{cp}}(\sqrt{3}/2) = 2$ for the range of $\epsilon$ considered.

Finally, we define $ P_\| $ and $ P_\perp $ as the longitudinal and transverse pressure components. Their average gives the total pressure,
\begin{equation}
	P = \frac{P_\| + 2P_\perp}{3}.
\end{equation}
Similarly to the linear density, it is also useful here to introduce a one-dimensional version of the longitudinal pressure,
\begin{equation}
	p = \frac{\pi \epsilon^2}{4} P_\|.
\end{equation}

\subsection{\label{sec:laplacetransform}Radial distribution function}

The RDF in a Q1D system quantifies positional correlations as a function of the longitudinal separation $x$. Let us denote as $\phi_\rr^2$ the composition distribution function, so that $\phi_{\mathbf{r}}^2 d^2\mathbf{r}$ is the fraction of particles in the system that are located at an elementary transverse position $d^2\rr$ in the cylinder's cross section. The normalization condition then becomes
\begin{equation}
\label{eq:normalization_1}
\int d^2\rr\,\phi^2_\rr=\pi\int_0^{\epsilon^2/4} du\,\phi^2_u=1,
\end{equation}
where the notation \(u \equiv r^2\) and \(\phi_\rr \to \phi_u\) has been introduced for mathematical convenience.

In the isothermal--isobaric ensemble at fixed reduced pressure $\beta p$, where $\beta = 1/k_B T$ is the inverse temperature and $k_B$ is the Boltzmann constant, the exact theoretical treatment is built from the first nearest-neighbor probability distribution~\cite{S16,MS25}
\begin{equation}
	\label{PP1_rew}
	\PP^{(1)}_{\rr_1,\rr_2}(x)=\frac{\beta p}{\ell}\,
	\frac{\phi_{u_2}}{\phi_{u_1}}\,
	e^{-\beta p x}\,\Theta\!\left(x-a_{\rr_1,\rr_2}\right),
\end{equation}
where $\Theta\!\left(\cdot\right)$ is the Heaviside step function. Equation~\eqref{PP1_rew} represents the probability density that the first nearest-neighbor of a particle at a transverse position $\rr_1=(u_1,\theta_1)$ is located at a transverse position $\rr_2=(u_2,\theta_2)$ and at a longitudinal distance $x$.
 
In this framework, both $\ell$ and $\phi_u^2$ are obtained simultaneously by solving the integral eigenvalue equation~\cite{KP93,MS25}
\begin{equation}
	\label{eq:eigenfunction2_rew}
	\frac{1}{2}\int_0^{\epsilon^2/4}\!du_2\,\phi_{u_2}\int_0^{2\pi}\!d\theta_{12}\,
	e^{-\beta p\,a_{\rr_1,\rr_2}}
	=\ell\,\phi_{u_1}
\end{equation}
and retaining its largest eigenvalue and corresponding eigenvector. The linear density of the system is then directly obtained as~\cite{KP93,MS25}
\begin{equation}
	\frac{1}{\lambda} = \frac{1}{\beta p} - \frac{\partial \ln \ell}{\partial( \beta p)}.
\end{equation}

 Higher-order nearest-neighbor distributions follow from the convolution hierarchy
\begin{align}
	\label{eq:convP1_rew}
	\PP^{(n)}_{\rr_1,\rr_2}(x)=&\frac{1}{2}\int_0^{\epsilon^2/4}\!du_3\int_0^{2\pi}\!d\theta_{13}\int_0^x\!dx' \PP^{(n-1)}_{\rr_1,\rr_3}(x')\, \nonumber \\
	&\times \PP^{(1)}_{\rr_3,\rr_2}(x-x').
\end{align}
The transverse-resolved RDF is then written as the sum over neighbor orders,
\begin{equation}
	\label{eq:gpartial_rew}
	g_{\rr_1,\rr_2}(x)=\frac{1}{\lambda\,\phi^2_{u_2}}\sum_{n=1}^\infty \PP^{(n)}_{\rr_1,\rr_2}(x),
\end{equation}
while the overall RDF is obtained by averaging over all transverse configurations,
\begin{equation}
	\label{eq:gtotal_rew}
	g(x)=\frac{\pi}{2}\int_0^{\epsilon^2/4}\!du_1\,\phi^2_{u_1}\int_0^{\epsilon^2/4}\!du_2\,\phi^2_{u_2}
	\int_0^{2\pi}\!d\theta_{12}\;g_{\rr_1,\rr_2}(x).
\end{equation}

Because of the convolution structure of $\PP^{(n)}_{\rr_1,\rr_2}(x)$ in Eq.~\eqref{eq:convP1_rew}, it proves advantageous to formulate the RDFs in Laplace space.  Defining $\widetilde{\mathcal{P}}_{\rr_1,\rr_2}^{(1)}(s)=\mathcal{L}\{\mathcal{P}_{\rr_1,\rr_2}^{(1)}(x)\}$, where $\mathcal{L}\{\cdot\}$ is the Laplace-transform operator, one finds
\begin{equation}
	\label{eq:P1_laplace_angular}
	\widetilde{\mathcal{P}}_{\rr_1,\rr_2}^{(1)}(s)
	=
	\frac{\beta p}{\left(s+\beta p\right) \ell}
	\frac{\phi_{u_2}}{\phi_{u_1}}
	e^{
		-\left(s+\beta p\right)
		a_{\rr_1,\rr_2}}.
\end{equation}

The Laplace transform of the RDF $\widetilde{G}_{\rr_1,\rr_2}(s)=\mathcal{L}\{g_{\rr_1,\rr_2}(x)\}$ is then~\cite{MS25}
\begin{align}
	\label{eq:Gs01_rew}
	\widetilde{G}_{\rr_1,\rr_2}(s)=&
	\frac{1}{\lambda\,\phi^2_{u_2}} 	\left(\sum_{n=1}^{\infty}\left[\widetilde{\mathsf{P}}^{(1)}(s)\right]^n\right)_{\rr_1,\rr_2} \nonumber \\
	=&\frac{1}{\lambda\,\phi^2_{u_2}}
	\left(\widetilde{\mathsf{P}}^{(1)}(s)\cdot\left[\mathsf{I}-\widetilde{\mathsf{P}}^{(1)}(s)\right]^{-1}\right)_{\rr_1,\rr_2},
\end{align}
where $\mathsf{I}$ is the identity operator and $\widetilde{\mathsf{P}}^{(1)}(s)$ is the operator with kernel $\widetilde{\mathcal{P}}_{\rr_1,\rr_2}^{(1)}(s)$.

The Laplace transform $\widetilde{G}_{\rr_1,\rr_2}(s)$ depends on the angular variables only through $\cos\theta_{12}$ inside $a_{\rr_1,\rr_2}$ and it is therefore invariant under rotations around the pore axis ($\theta_1 \to \theta_1+\alpha$ and $\theta_2 \to \theta_2+\alpha$) and under reflection ($\theta_1 \to -\theta_1$ and $\theta_2 \to -\theta_2$), which means that it admits a cosine Fourier expansion of the form
\begin{equation}
	\label{eq:G_cosine_expansion}
	\widetilde{G}_{\rr_1,\rr_2}(s)
	= \frac{Q_{u_1,u_2}^{(0)}(s)}{2}
	+
	\sum_{m=1}^{\infty}
	Q_{u_1,u_2}^{(m)}(s)
	\cos(m\theta_{12}),
\end{equation}
where
\begin{equation}
	\label{eq:G_cosine_coefficients}
	Q_{u_1,u_2}^{(m)}(s)
	=
	\frac{2}{\pi}
	\int_0^{\pi}d\theta_{12}\,
\widetilde{G}_{\rr_1,\rr_2}(s)
	\cos(m\theta_{12}).
\end{equation}

As for the overall RDF, the Laplace transform of Eq.~\eqref{eq:gtotal_rew} is
\begin{equation}
	\label{eq:gtotalLaplace_rew}
	\widetilde{G}(s)=\frac{\pi}{2}\int_0^{\epsilon^2/4}\!du_1\,\phi^2_{u_1}\int_0^{\epsilon^2/4}\!du_2\,\phi^2_{u_2}
	\int_0^{2\pi}\!d\theta_{12}\;\widetilde{G}_{\rr_1,\rr_2}(s).
\end{equation}
Inserting Eq.~\eqref{eq:G_cosine_expansion} into Eq.~\eqref{eq:gtotalLaplace_rew} yields
	 \begin{align}\label{eq:global_G_only_m0}
	\widetilde{G}(s)
	=&
	\frac{\pi^2}{2}
	\int_0^{\epsilon^2/4}du_1\,\phi_{u_1}^{2}
	\int_0^{\epsilon^2/4}du_2\,\phi_{u_2}^{2}
	Q_{u_1,u_2}^{(0)}(s),
\end{align}
which means that only $Q_{u_1,u_2}^{(0)}(s)$ enters as an active function inside the global RDF, since all other terms with $m\geq1$ cancel out upon angular averaging.

Although the calculations are performed exactly in Laplace space, the analysis of correlations sometimes requires recovering $g_{\rr_1,\rr_2}(x)$ and $g(x)$ in real space. In these cases, the inverse Laplace transform of Eqs.~\eqref{eq:Gs01_rew} and~\eqref{eq:gtotalLaplace_rew} can be done analytically for the simplest potentials~\cite{S16} or by means of numerical algorithms.~\cite{AW92}

\subsection{Correlation lengths}

The correlation length, $\xi$, characterizes the spatial extent over which structural correlations persist within a system, providing a quantitative measure of how rapidly the positional correlations between particles decay with distance. It is commonly determined from the asymptotic behavior of the RDF or, equivalently, from the correlation functions $h_{\rr_1,\rr_2}(x)=g_{\rr_1,\rr_2}(x)-1$ and $h(x)=g(x)-1$. 

\subsubsection{Laplace-space representation}
The Laplace-space representation of the RDFs given in Eqs.~\eqref{eq:Gs01_rew} and~\eqref{eq:gtotalLaplace_rew} enables a direct analysis of the asymptotic behavior of the correlation function $h_{\rr_1,\rr_2}(x)$, which is controlled by the singularity structure of $\widetilde{G}_{\rr_1,\rr_2}(s)$. In the limit of large separations $x$, the long-range decay of $h_{\rr_1,\rr_2}(x)$ is governed primarily by the pole with the largest real part---the pole closest to the imaginary axis---since higher-order contributions decay more rapidly.

The correlation function in the long-range asymptotic regime can then be approximated by the contribution of this leading pole, which we denote by $\smin$. Note that, throughout the remainder of this work, we use the term ``pole'' generically to denote either an isolated real pole or a complex-conjugate pair.

If the leading pole is real, $\smin=-\kappa$, the asymptotic decay is a monotonically decaying exponential
\begin{equation}
	\label{eq:realpole_rew}
	h_{\rr_1,\rr_2}(x)\approx \mathcal{A}_{\rr_1,\rr_2}^{(\min)}\,e^{-\kappa x},
\end{equation}
with the residue $\mathcal{A}_{\rr_1,\rr_2}^{(\min)}$ being a real number. On the other hand, if the leading singularities form a complex conjugate pair, $\smin=-\kappa\pm \imath \omega$, then grouping both residues gives a damped oscillatory decay,
\begin{equation}
	\label{eq:complexpole_rew}
	h_{\rr_1,\rr_2}(x)\approx 2\,\left|\mathcal{A}_{\rr_1,\rr_2}^{(\min)}\right|\,e^{-\kappa x}\cos(\omega x+\delta_{\rr_1,\rr_2}^{(\min)}),
\end{equation}
with $\mathcal{A}_{\rr_1,\rr_2}^{(\min)}=\left|\mathcal{A}_{\rr_1,\rr_2}^{(\min)}\right| e^{\imath \delta^{(\min)}_{\rr_1,\rr_2} }$ being a complex number.

In both cases, irrespective of whether the asymptotic behavior is monotonic or oscillatory, the decay rate $\kappa$ is directly associated with the correlation length
\begin{equation}
	\xi = \kappa^{-1}.
\end{equation}

\subsubsection{Selection by residue symmetry}

 Following the decomposition in Eq.~\eqref{eq:G_cosine_expansion}, the set of poles of $\widetilde{G}_{\rr_1,\rr_2}(s)$ is the union of all the $s_n^{(m)}$ poles of each Fourier coefficient $Q_{u_1,u_2}^{(m)}(s)$. Defining
\begin{equation}
	A_{u_1,u_2}^{(n,m)}=\operatorname*{Res}_{s=s_n^{(m)}}	\left[ Q_{u_1,u_2}^{(m)}(s) \right],
\end{equation}
it follows that the residue of the full transverse-resolved RDF is
\begin{equation}
	\label{eq:A_from_Gm_residues}
	\mathcal{A}_{\rr_1,\rr_2}^{(n,m)}=\begin{cases}
		\dfrac{1}{2}A^{(n,0)}_{u_1,u_2}, & m=0,\\
		A^{(n,m)}_{u_1,u_2} \cos(m\theta_{12}), & m\geq1,
	\end{cases}
\end{equation}
which shows that the angular dependence of each residue is directly inherited from the angular dependence of the corresponding  $Q_{u_1,u_2}^{(m)}(s)$.  The transverse correlation functions can be reconstructed from the poles in Eq.~\eqref{eq:A_from_Gm_residues} as
\begin{align}\label{eq:h12_frompoles2}
	h_{\rr_1,\rr_2}(x) = &\frac{1}{2}\sum_n e^{s_n^{(0)} x}\,A_{u_1,u_2}^{(n,0)} \nonumber \\&+  \sum_{m\geq1} \cos(m \theta_{12}) \sum_n  e^{s_n^{(m)} x}\,A_{u_1,u_2}^{(n,m)}.
\end{align}

An interesting consequence of Eqs.~\eqref{eq:A_from_Gm_residues} and~\eqref{eq:h12_frompoles2} is that different transverse configurations can have very different contributions from different poles, due to vanishing residues. Whenever $\cos(m\theta_{12})=0$, the corresponding residue  $\mathcal{A}_{\rr_1,\rr_2}^{(n,m)}$ vanishes and the pole does not contribute to that particular transverse-resolved RDF, even though it remains present in the full angular-dependent function. For $m\geq 1$, these angles are
\begin{equation}
	\label{eq:zero_angles_correct}
\theta_{12}
	=
	\frac{(2k+1)\pi}{2m},
	\qquad
	k=0,1,\ldots,2m-1.
\end{equation}

Finally, let us analyze the contributions to the global correlation function $h(x)$, which is given by the weighted average of all transverse-resolved ones. Taking into account Eq.~\eqref{eq:gtotal_rew} and the decomposition in Eq.~\eqref{eq:h12_frompoles2}, one finds
\begin{equation}\label{eq:h_frompoles2}
	h(x) =\sum_{n} e^{s_n^{(0)} x} 
	\mathcal{A}^{(n,0)},
\end{equation}
where
\begin{equation}
	\label{eq:Atotal2}
	\mathcal{A}^{(n,0)}= \frac{\pi^2}{2}\int_0^{\epsilon^2/4}\!du_1\,\phi^2_{u_1}\int_0^{\epsilon^2/4}\!du_2\,\phi^2_{u_2}
	A_{u_1,u_2}^{(n,0)}.
\end{equation}
Here we have taken into account that the angular integration makes all terms with $m\geq1$ vanish. Consequently, the poles associated with a mode $m\geq1$ do not contribute to the global RDF. This analysis can also be extracted from Eq.~\eqref{eq:global_G_only_m0}, since only the term $Q_{u_1,u_2}^{(0)}(s)$ contributes to the global $\widetilde{G}(s)$.

\section{Numerical details}\label{sec:numerics}

The pole structure of $\widetilde{G}_{\rr_1,\rr_2}(s)$ determines the asymptotic behavior of the longitudinal pair correlations and, therefore, the corresponding correlation lengths. However, locating the leading poles accurately is numerically nontrivial, particularly
in regions where different poles have similar decay rates or where a pole has a vanishing residue for a particular transverse-resolved correlation function.

Although the angular decomposition introduced in Eq.~\eqref{eq:G_cosine_expansion} is useful for interpreting the symmetry of the residues, the numerical calculations were performed
by discretizing the full transverse operator $\widetilde{\mathsf P}^{(1)}(s)$ in the variables
$\rr=(u,\theta)$. We introduced an equidistant radial $\{u_i\}_{i=1}^{N_u}$ and angular $\{\theta_\alpha\}_{\alpha=1}^{N_\theta}$ grid, and denote the resulting transverse grid points by
\begin{equation}
	\rr_{i\alpha}=(u_i,\theta_\alpha),
	\qquad
	i=1,\ldots,N_u,
	\qquad
	\alpha=1,\ldots,N_\theta.
\end{equation}
The continuous operator $\widetilde{\mathsf{P}}^{(1)}(s)$ is then approximated by the matrix $\widetilde{\mathbf P}^{(1)}(s)$,
\begin{equation}
	\left(	\widetilde{\mathbf P}^{(1)}(s)
	\right)_{i\alpha,j\beta}
	= \frac{\pi \epsilon^2}{4 N_u N_\theta}
	\widetilde{\mathcal P}_{\rr_{i\alpha},\rr_{j\beta}}^{(1)}(s).
	\label{eq:P_full_matrix}
\end{equation}
Thus, for each complex value of $s$, the full transverse operator is
represented by an
$(N_uN_\theta)\times(N_uN_\theta)$ matrix. Within this discretized representation, following Eq.~\eqref{eq:Gs01_rew}, the poles are obtained as the complex roots of
\begin{equation}
	\label{eq:poles_det_rew}
	\det\!\left[
	\mathbf I-\widetilde{\mathbf P}^{(1)}(s)
	\right]
	=
	0.
\end{equation}
Because the physical parameters entering the kernel are real, complex
poles occur in conjugate pairs. Hence, it is sufficient to locate the
poles in the upper half of the complex $s$ plane and to include their
complex conjugates when reconstructing real-space correlation
functions.

The discretized analogue of Eq.~\eqref{eq:Gs01_rew} is obtained from
\begin{equation}
	\widetilde{G}_{\rr_{i\alpha},\rr_{j\beta}}(s)
	\simeq
	\frac{4 N_u N_\theta}{\lambda\phi_{u_j}^{2} \pi \epsilon^2}
	\left(
	\widetilde{\mathbf P}^{(1)}(s)
	\cdot
	\left[
	\mathbf I-\widetilde{\mathbf P}^{(1)}(s)
	\right]^{-1}
	\right)_{i\alpha,j\beta}.
	\label{eq:G_discrete_reconstruction}
\end{equation}

Once a pole $s_k$ has been located, its residue in a particular
transverse-resolved RDF is evaluated using
\begin{equation}
	\mathcal A_{\rr_1,\rr_2}^{(k)}
	=
	\left[
	\left.
	\frac{d}{ds}
	\left(
	\widetilde{G}_{\rr_1,\rr_2}(s)^{-1}
	\right)
	\right|_{s=s_k}
	\right]^{-1}.
	\label{eq:residue_inverse_derivative}
\end{equation}

The numerical calculations were repeated for several values of the radial and angular resolutions, $N_u$ and $N_\theta$. For the range of pressures investigated here, resolutions of order $N_u\simeq 50$ and $N_\theta\simeq 50$ were typically sufficient to obtain stable finite-grid estimates. The final reported values were obtained by extrapolating the finite-resolution results to the continuum-grid limit $N_u,N_\theta\to\infty$. Convergence was monitored independently for the real and imaginary parts of the leading poles, for the associated residues, and for the resulting correlation lengths. Special care was taken near pole crossings, where small numerical uncertainties in the pole locations can affect the identification of the dominant asymptotic contribution.

\section{Results}\label{sec:results}

To illustrate the results, we are going to focus on a system with $\epsilon=0.866\simeq \sqrt{3}/2$, which is the maximum width of the channel that still allows for an exact solution using the framework described in Sec.~\ref{sec:theory}.

\subsection{Analysis of the leading poles}\label{sec:results_leadingpoles}

\begin{figure}
	\includegraphics[width=\columnwidth]{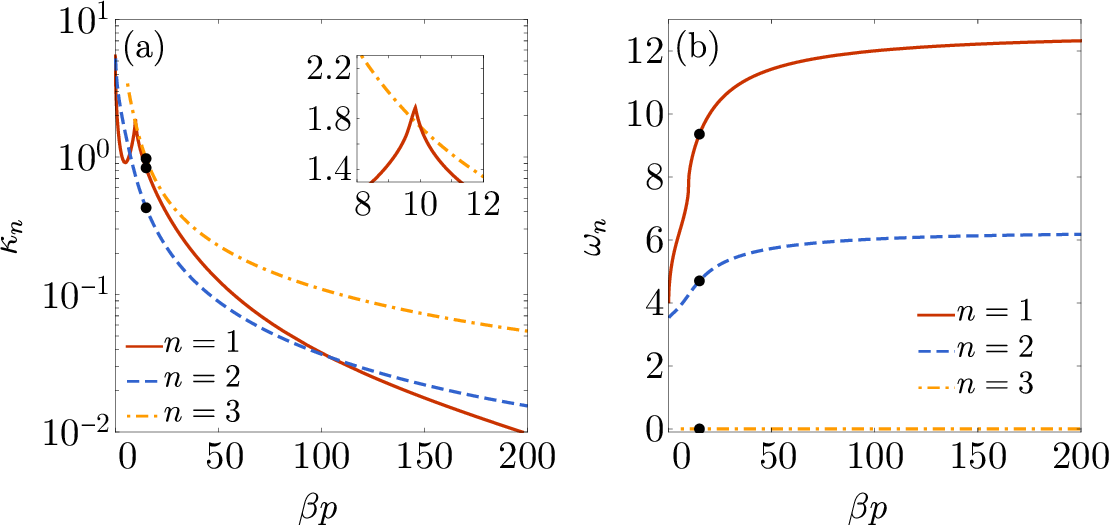}
	\caption{(a) Real and (b) imaginary parts of the three most relevant poles for a channel of width $\epsilon = 0.866$. The inset magnifies the region $8 < \beta p < 12$. Dots indicate the pole values at $\beta p = 15$.}
	\label{fig:poles}
\end{figure}

A representative subset of the pole spectrum for confined hard spheres in a channel of width $\epsilon = 0.866$ is shown in Fig.~\ref{fig:poles}, where panels (a) and (b) display the real and imaginary parts of the poles, respectively. The poles, denoted by $s_1$, $s_2$, and $s_3$, are labeled according to the ordering of their decay rates $\kappa_n$ at low pressure, and this convention is maintained across the full pressure range for clarity.

As expected, the decay rates $\kappa_n$ generally decrease with increasing pressure, although the behavior is not strictly monotonic. At low pressures, the ordering $\kappa_1 < \kappa_2 < \kappa_3$ holds, but this hierarchy is not preserved at intermediate pressures, where several crossings occur between the poles. These crossings imply changes in the dominant pole, although, depending on the vanishing residues and symmetry properties, they do not necessarily translate into changes in the effective leading contribution.

Of particular interest is the kink in $s_1$ near $\beta p \simeq 9.7$. In a very narrow pressure interval around this point, we find $\kappa_3<\kappa_1$, indicating a crossing between the two poles. Although the interval is small and the difference between the decay rates is slight, a detailed numerical analysis with increased resolution consistently supports the existence of both the crossing and the associated kink.

Of particular interest is the region around $\beta p \simeq 10$, where $\kappa_1$ and $\kappa_3$ become nearly degenerate. In this regime, the corresponding decay rates are very similar, indicating that both poles contribute to the correlation functions over a wide range of distances. As a result, the onset of the asymptotic regime is significantly delayed, since the dominance of a single pole emerges only at sufficiently large separations.

In contrast with the general behavior of $\kappa_n$, the oscillation frequencies $\omega_1$ and $\omega_2$ saturate at relatively low pressures, approaching quickly their asymptotic values $\omega_1 \simeq 4\pi$ and $\omega_2 \simeq 2\pi$, respectively. This indicates that the characteristic asymptotic wavelength of the correlations is established early, while the decay length continues to evolve. The pole $s_3$ is purely real, and therefore $\omega_3 = 0$ for all pressures.

\subsection{Residue contributions}

The asymptotic behavior and correlation length of the standard RDF $g(x)$ and the corresponding poles of $\widetilde{G}(s)$ will be analyzed in detail. However, the large number of possible transverse-resolved RDFs $g_{\rr_1,\rr_2}(x)$ makes a complete exploration of partial RDFs impractical. We therefore restrict our attention to the most relevant configurations. Specifically, we consider pairs of peripheral particles---those in contact with the confining walls, where $r_1 = r_2 = \epsilon/2$---and focus on a representative set of relative orientations given by $\theta_{12} = 0, \pi/2, \pi$.

\begin{figure}
	\includegraphics[width=\columnwidth]{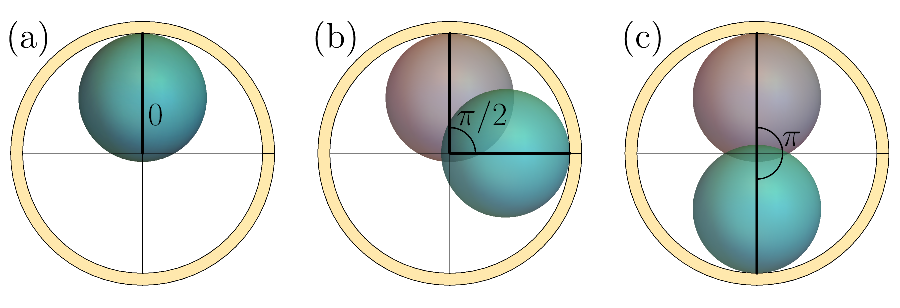}
	\caption{Schematic representation of the relative positions of the peripheral particles for which the transverse-resolved RDF will be analyzed: (a) $g_0(x)$, (b) $g_{\pi/2}(x)$ and (c) $g_\pi(x)$.}
	\label{fig:positions}
\end{figure}

For notational convenience, we introduce the shorthand
\begin{equation}
	\label{gtheta}
	g_{\theta}(x)\equiv \left.g_{\rr_1,\rr_2}(x)\right|_{r_1=r_2=\frac{\epsilon}{2}},\quad \theta=\theta_{12},
\end{equation}
and adopt the same convention for any other quantity carrying analogous subscripts, such as $\mathcal{A}_{\rr_1,\rr_2}^{(n)}$. A schematic illustration of these configurations is provided in Fig.~\ref{fig:positions}.

\begin{figure}
	\includegraphics[width=\columnwidth]{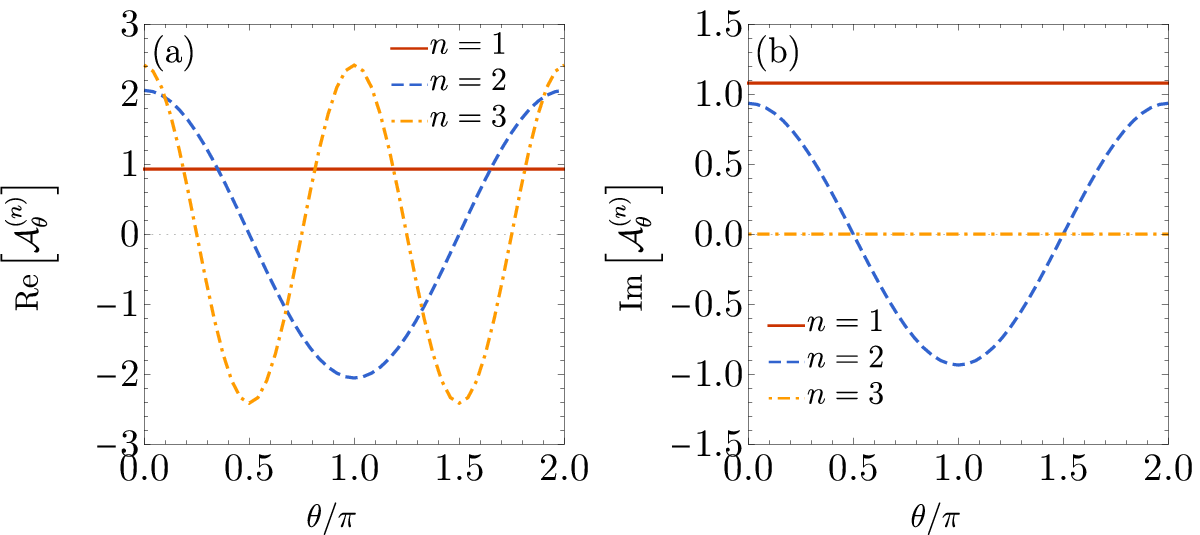}
	\caption{(a) Real and (b) imaginary parts of the corresponding residues as functions of the relative angle between peripheral particles at $\beta p = 15$.}
	\label{fig:residues}
\end{figure}

An insightful way to assess the role of the three poles $s_1$, $s_2$, and $s_3$ is to examine how their residues $\mathcal{A}_{\rr_1,\rr_2}^{(n)}$ depend on orientation while keeping the radial coordinates fixed. As a representative example, Fig.~\ref{fig:residues} shows the real and imaginary parts of the residues $\mathcal{A}^{(n)}_{\theta}$ of peripheral particles at a pressure $\beta p = 15$.

Although the poles are obtained numerically from the full transverse-resolved RDF, without resolving the operator into angular components, their angular character can be identified from the angular dependence of the corresponding residues. For the three leading poles considered here, the numerical residues are found to be of the form given by Eq.~\eqref{eq:A_from_Gm_residues}, with
 \begin{equation} m_1=0, \qquad m_2=1, \qquad m_3=2. \label{eq:leading_pole_angular_classes} 
 \end{equation}
 Thus, $s_1$ has an angle-independent residue and therefore will contribute to all channels, whereas $s_2$ and $s_3$ display different angular dependences which determine which poles contribute to each transverse-resolved RDF. In particular,  $s_2$ contributes to all channels with different amplitudes, except for $\theta=\pi/2$ and $\theta=3\pi/2$, where the residue vanishes. In the case of $s_3$, it contributes to all configurations except for $\theta=\pi/4$, $\theta=3\pi/4$, $\theta=5\pi/4$, and $\theta=7\pi/4$.
 
 The same classification also explains the relation between the transverse-resolved and global RDFs. Since the global RDF involves an average over the relative angle, the contributions from $s_2$ and $s_3$, for which $m_n\geq 1$, vanish upon angular integration,  and, as a result, they do not contribute to the global RDF. Consequently, among the three poles considered here, only pole $s_1$ contributes to the global RDF, while the poles identified as $s_2$ and $s_3$ can only be detected through transverse-resolved correlation functions.

As a summary of all the above, Table.~\ref{tab:pole_contributions} shows which of the three considered poles contributes to which RDF.

\begin{table}[h!]
	\centering
	\renewcommand{\arraystretch}{1.35} 
	\setlength{\tabcolsep}{12pt} 
	\begin{tabular}{lccc}
		\hline
		& $s_1$ & $s_2$ & $s_3$ \\
		\hline
		$g(x)$        & \checkmark & $-$ & $-$ \\
		$g_{0}(x)$       & \checkmark & \checkmark & \checkmark \\
		$g_{\pi/2}(x)$   & \checkmark & $-$ & \checkmark \\
		$g_{\pi}(x)$     & \checkmark & \checkmark & \checkmark \\
		\hline
	\end{tabular}
	\caption{Contribution of the poles $s_1$, $s_2$, and $s_3$ to the global and transverse-resolved RDFs.}
	\label{tab:pole_contributions}
\end{table}

It should be emphasized that the poles $s_1$, $s_2$, and $s_3$ do not necessarily correspond, at a given pressure, to those nearest to the real axis. Instead, as said before, they have been selected because each one of them yields the dominant contribution to at least one of the considered transverse-resolved functions $\widetilde{G}_{\rr_1,\rr_2}(s)$ at a certain pressure within the range considered. This makes them the most relevant poles for the present analysis, even if other poles may lie between them in the full spectrum.

\subsection{Different decays for different RDFs}

The analysis of the dominant poles and their associated residues provides direct access to the long-distance asymptotic behavior of the RDFs, including both their decay rate and oscillation frequency. Due to symmetry-induced cancellations, different RDFs—particularly the transverse-resolved ones—can exhibit markedly different correlation lengths. As a result, specific transverse channels may display significantly more structured behavior (i.e., larger correlation lengths) than others, and even larger than the global RDF.

\begin{figure}
	\includegraphics[width=\columnwidth]{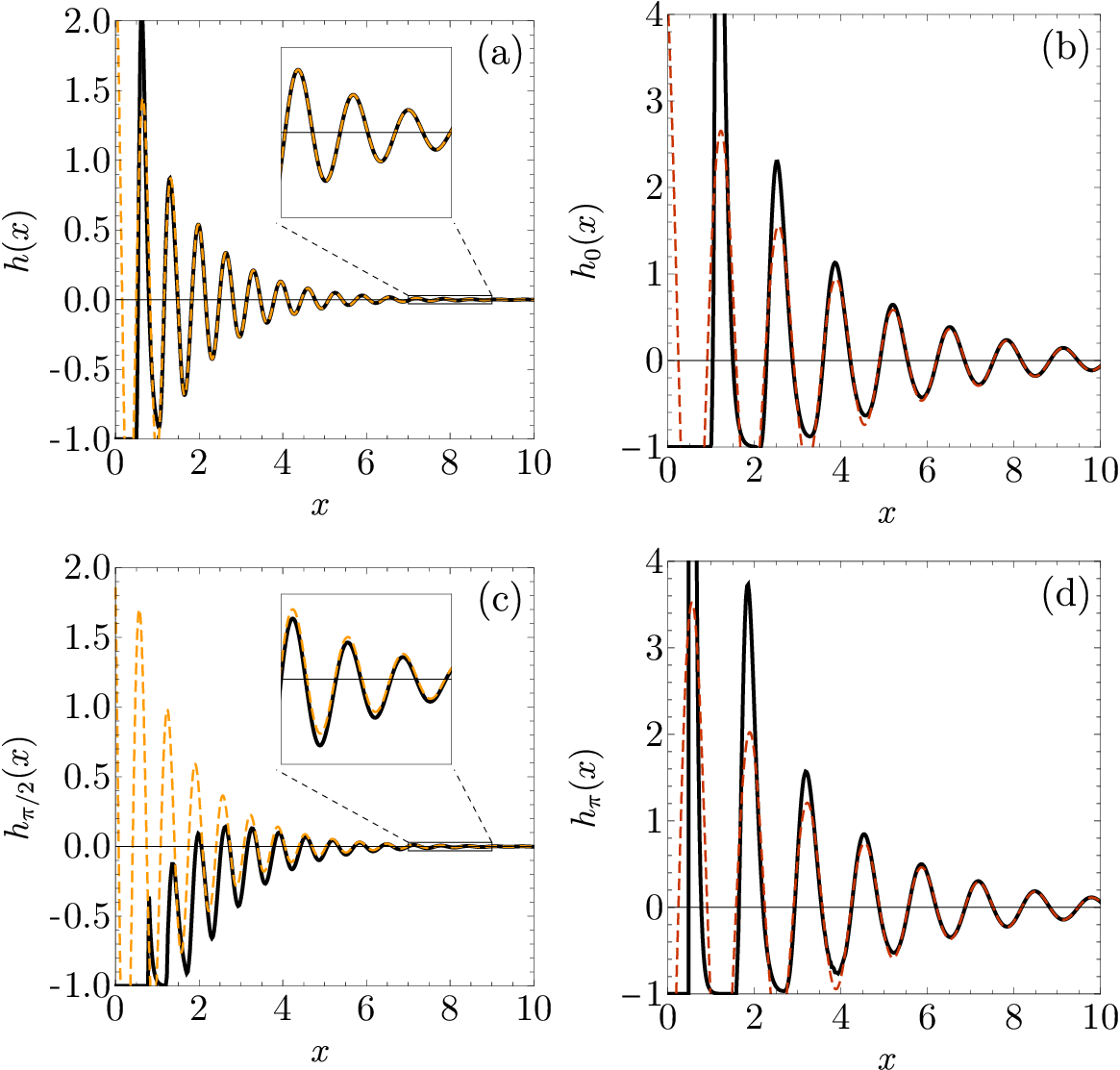}
	\caption{Correlation functions for the global and transverse-resolved RDFs (solid lines), compared with their asymptotic forms obtained from the pole structure by retaining only the dominant pole (dashed lines).}
	\label{fig:gxreconstruct}
\end{figure}

Figure~\ref{fig:gxreconstruct} shows several correlation functions at a representative value of $\beta p = 15$, together with their long-distance approximations [Eqs.~\eqref{eq:h12_frompoles2} and~\eqref{eq:h_frompoles2}], constructed from the corresponding leading pole. The global function $h(x)$, shown in Fig.~\ref{fig:gxreconstruct}(a), is already very well described by its asymptotic form at relatively short distances, indicating that no other relevant poles with comparable values of $\kappa$ contribute significantly at that pressure.

A different behavior is observed for the transverse-resolved functions $h_{0}(x)$ and $h_{\pi}(x)$, shown in Figs.~\ref{fig:gxreconstruct}(b) and~\ref{fig:gxreconstruct}(d), respectively. In these cases, the leading-pole approximation becomes accurate only at somewhat larger distances. These two functions also display decay rates and oscillation frequencies that clearly differ from those of the global correlation function.

The case of $h_{\pi/2}(x)$, displayed in Fig.~\ref{fig:gxreconstruct}(c), is particularly interesting. Although its effective leading pole is the same as that of the global $h(x)$, the simultaneous contribution of both $s_1$ and $s_3$, with very similar values of $\kappa_1 \simeq \kappa_3$ and comparable amplitudes, delays the onset of the asymptotic regime. As a result, even at $x \simeq 8$ the correlation function is not yet well captured by the leading-pole approximation alone, and a more accurate reconstruction at those distances would require the inclusion of the subleading pole $s_3$.

\begin{figure}
	\includegraphics[width=\columnwidth]{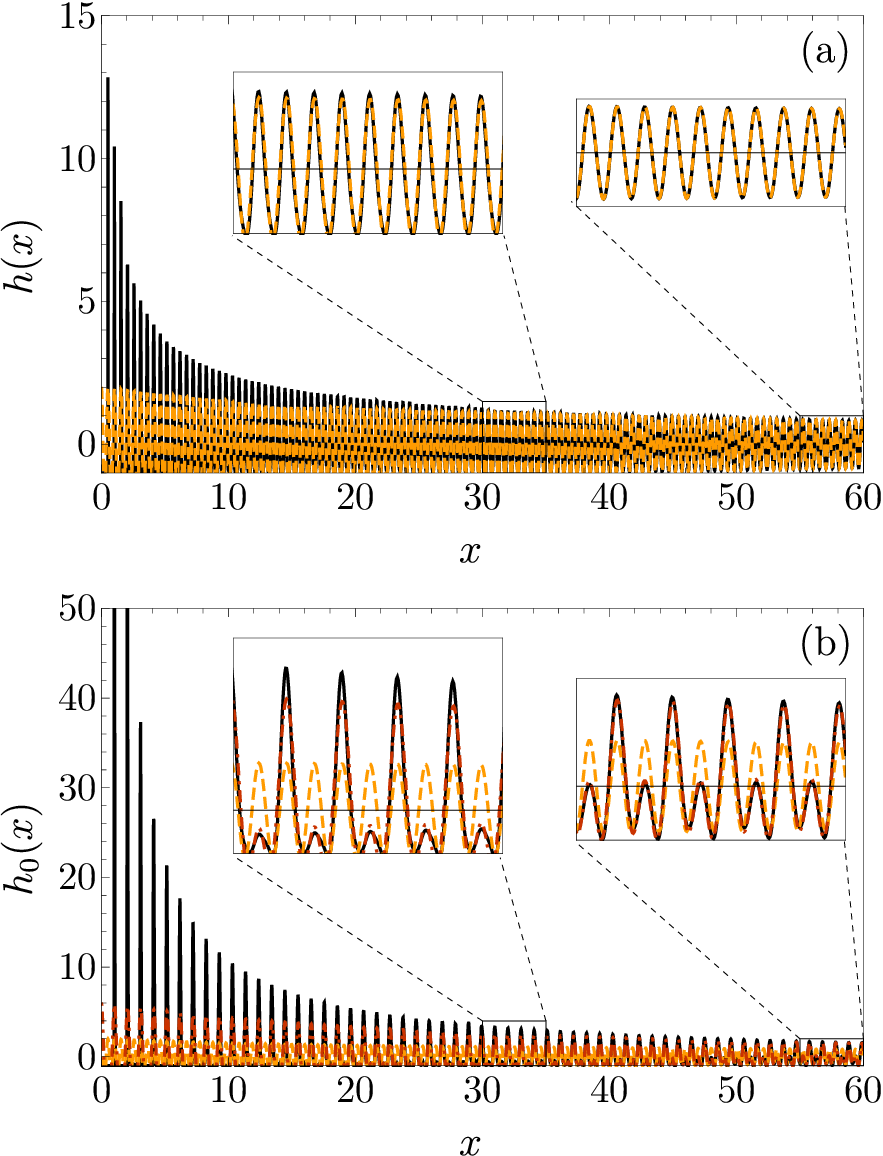}
	\caption{Correlation functions (a) $h(x)$ and (b) $h_0(x)$ at $\beta p=100$ (solid lines), compared with their asymptotic forms obtained from the pole structure by retaining only the dominant pole (dashed lines) and, in panel (b), also the first two dominant poles (dashed-dotted line).}
	\label{fig:GxLongDistance}
\end{figure}

To better illustrate the competition between poles and the role of subleading contributions, Fig.~\ref{fig:GxLongDistance} shows $h(x)$ and $h_0(x)$ at $\beta p = 100$. Although both RDFs share the same leading pole at this pressure, the rate at which they approach the asymptotic behavior is markedly different. This difference arises from the influence of subleading poles, which remain significant over intermediate distances. In this particular scenario, the leading pole at $\beta p=100$ is $s_1$, but the subleading pole $s_2$ has a very similar decay rate [see Fig.~\ref{fig:poles}]. 

This proximity between both poles is not relevant in the case of $h(x)$, since only $s_1$ contributes and the asymptotic regime sets in relatively fast and is well captured by the contribution of a single pole. In fact, at intermediate distances ($x \simeq 40$) the leading-pole approximation already reproduces the oscillation frequency very accurately, even though it still underestimates the peak heights.

The oscillatory structure of $h(x)$ sets in so rapidly that at distances as close as $x \simeq 5$ the function already oscillates with nearly the same frequency $\omega_1 \simeq 4\pi$ that characterizes its large-$x$ behavior. On the other hand, the decay rate does not match the asymptotic one until distances of around $x \simeq 55$, from where it also provides an accurate description of the oscillation amplitude. This behavior is consistent with the fact that $h(x)$ receives contributions only from the pole $s_1$, and the rest of the poles that might contribute (not shown in Fig.~\ref{fig:poles}) have much larger decay rates.

The behavior of $h_0(x)$ is considerably more complex because of the additional influence and proximity of $s_2$. As seen in Fig.~\ref{fig:poles}(a), at $\beta p = 100$ the decay rates associated with these two poles, $\kappa_1$ and $\kappa_2$, are very similar. As a result, although the ultimate asymptotic decay is still governed by $s_1$, the contribution from $s_2$ remains relevant over a wide range of distances, which delays the onset of the true asymptotic regime.

Figure~\ref{fig:GxLongDistance}(b) shows that the behavior of $h_0(x)$ is much more complex over the full distance range. Unlike the global correlation function, the short-distance oscillations of $h_0(x)$ differ significantly from the asymptotic ones. Moreover, even at relatively large distances, $x \simeq 60$, the leading-pole approximation still provides a poor description: it fails to reproduce both the decay and the oscillation frequency accurately.

This slow convergence toward the asymptotic regime is induced by the competition between $\kappa_1$ and $\kappa_2$. In fact, the insets of Fig.~\ref{fig:GxLongDistance}(b) show the gradual emergence of a secondary peak that eventually becomes comparable in importance to the main one, and that cannot be captured by the inclusion of a single pole. At these distances, the contribution of the subleading pole therefore remains essential. Indeed, once both poles are included, the reconstruction of $h_0(x)$ becomes very accurate.

\subsection{Correlation lengths}

The behavior of the poles, together with the symmetry properties of their residues, leads to an important conclusion: there is no single correlation length that fully characterizes the system. Different transverse channels can exhibit markedly different correlation lengths, which may also differ from that of the global RDF.

This means that analyzing only the correlation length extracted from the global RDF can be misleading, as it does not necessarily correspond to the longest correlation in the system. In fact, certain transverse-resolved channels may sustain correlations over longer distances.

\begin{figure}
	\includegraphics[width=\columnwidth]{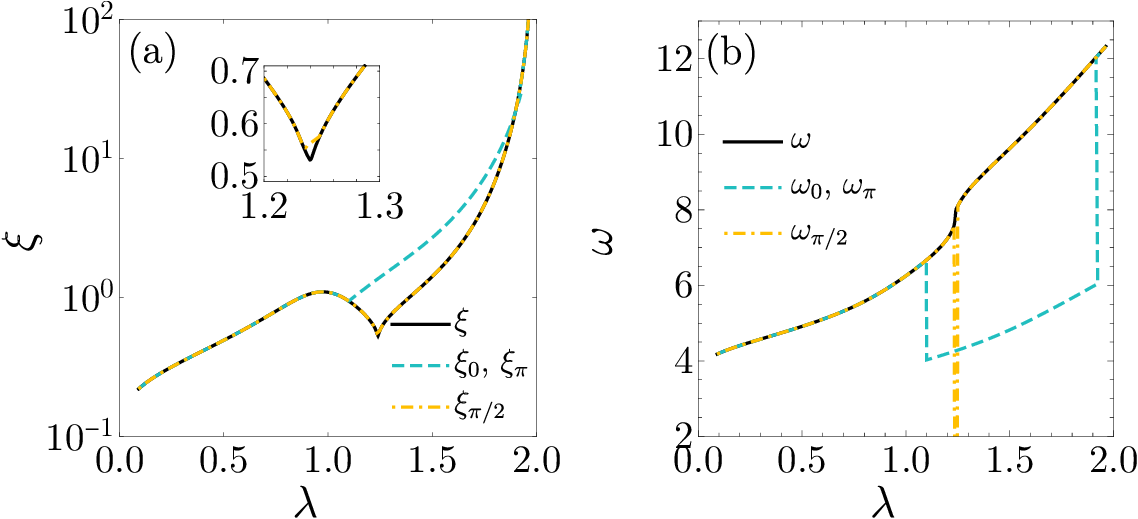}
	\caption{(a) Correlation lengths and (b) asymptotic oscillation frequencies of the global and transverse-resolved RDFs as functions of the linear density $\lambda$. The inset magnifies the region $1.2 < \lambda < 1.3$}
	\label{fig:correlations}
\end{figure}

Figure~\ref{fig:correlations}(a) shows the density dependence of the correlation lengths $\xi$ for the global correlation function $h(x)$, together with $\xi_\theta$ for the transverse-resolved functions with $\theta=0,\pi/2,$ and $\pi$. It highlights that different RDFs can exhibit distinct correlation lengths, particularly at intermediate densities. For clarity, the values of $\xi_0,\, \xi_{\pi}$, and $\omega_0,\,\omega_\pi$ are represented by common curves, since they are equal across the full pressure range.

At low densities, prior to the onset of the zigzag configuration, all RDFs share the same correlation length. As the density increases and the zigzag ordering develops (around a linear density $\lambda \approx 1$), correlations between peripheral particles located on the same and opposite sides of the channel become more persistent, leading to larger correlation lengths than that of the global RDF. This difference persists up to $\lambda \approx 1.9$, beyond which all correlation lengths converge again.

The correlation length $\xi_{\pi/2}$ closely follows that of the global $h(x)$, with the exception of a narrow region around $\lambda \simeq 1.25$, which follows the crossing between the poles shown the inset of in Fig.~\ref{fig:poles}(a). In this range, the global correlation length $\xi$ exhibits a sharp kink that is absent in $\xi_{\pi/2}$, leading to a small mismatch between the two. This difference is also reflected in the asymptotic oscillation frequencies shown in Fig.~\ref{fig:correlations}(b).

The behavior of these oscillation frequencies mirrors that of the correlation lengths. In particular, $\omega_0$ and $\omega_\pi$ deviate from the global frequency $\omega$ over the interval $\lambda \simeq 1$ to $\lambda \simeq 1.9$, whereas $\omega_{\pi/2}$ differs from $\omega$ only within a narrow region around $\lambda \simeq 1.25$. Notably, in this small interval around $\lambda \simeq 1.25$, the oscillation frequency $\omega_{\pi/2}$ vanishes, indicating that the transverse-resolved correlation functions with $\theta=\pi/2$ undergoes a purely monotonic decay in that narrow region.

\subsection{Behavior as the pore diameter decreases}

\begin{figure}
	\includegraphics[width=\columnwidth]{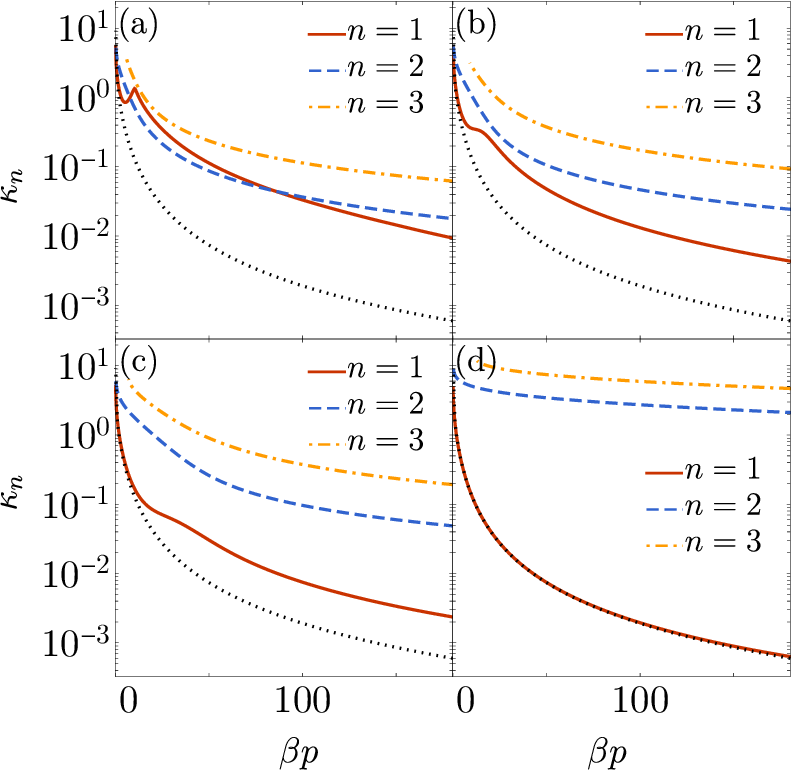}
	\caption{Real part of the poles as a function of pressure for (a) $\epsilon=0.85$, (b) $\epsilon=0.7$, (c) $\epsilon=0.5$, (d) $\epsilon=0.1$, along with the real part of the dominant pole of the Tonks gas (black, dotted line).}
	\label{fig:smalleps}
\end{figure}

Although the results have been analyzed primarily for $\epsilon = 0.866$, corresponding to the widest channel allowed within this exact framework, it is also instructive to examine how the pole structure evolves as the confinement is increased, i.e., as $\epsilon \to 0$. Figure~\ref{fig:smalleps} shows the evolution of the real part of the same three poles presented in Fig.~\ref{fig:poles}(a). As the channel narrows, $s_1$ rapidly becomes the dominant pole across the entire pressure range, already for values as large as $\epsilon \approx 0.7$. Since $s_1$ is the only pole that contributes to all RDFs, it follows that, for sufficiently small $\epsilon$, all RDFs share the same correlation length and asymptotic oscillation frequency.

Moreover, $s_1$ approaches the dominant pole of the Tonks gas in the limit $\epsilon \to 0$, as expected. In this limit, the system effectively reduces to a strictly one-dimensional fluid, and its behavior at finite pressure converges to that of the Tonks gas.

\section{Conclusions}\label{sec:conclusions}
In this work, we have presented a detailed analysis of the correlation lengths of hard spheres confined in a narrow cylindrical pore, focusing on the Q1D regime where only nearest-neighbor interactions are allowed. By exploiting the Laplace-space representation of the RDF, we have shown that the asymptotic decay of correlations can be directly obtained from the pole structure of the corresponding transform.

A central result of this study is that, although the pole spectrum is common to all correlation functions, the actual long-distance behavior depends critically on the associated residues. Symmetry-induced cancellations can suppress the contribution of the leading pole for specific transverse configurations, giving rise to effective leading poles that differ between transverse-resolved correlation functions.

This leads to the important conclusion that there is not a unique correlation length characterizing the system. In particular, the correlation length extracted from the global RDF may underestimate the true extent of correlations, since some transverse-resolved channels can sustain longer-ranged order that is not captured by the angularly averaged function. This effect is especially relevant at intermediate densities, where structural ordering associated with zigzag configurations enhances correlations between specific particle arrangements, as illustrated in Fig.~\ref{fig:correlations}.

We have also shown that the competition between poles can significantly delay the onset of the asymptotic regime, particularly in transverse-resolved correlations where multiple poles contribute with comparable decay rates. This highlights the importance of subleading contributions in describing the structure at intermediate distances.

Finally, we have examined the effect of increasing confinement. As the channel width decreases, a single pole progressively dominates the behavior of all correlation functions, and the system smoothly approaches the Tonks gas limit. In this regime, all correlation lengths and oscillation frequencies converge, recovering the expected one-dimensional behavior.

Overall, our results demonstrate that a transverse-resolved analysis is essential for a complete understanding of structural correlations in confined fluids. The approach developed here provides a general framework to uncover hidden correlations that are not accessible through global observables alone, and may be extended to more complex confined systems and interaction potentials.

\begin{acknowledgments}
	The author acknowledges financial support from Grant No.~PID2024-156352NB-I00 funded by MCIU/AEI/10.13039/501100011033 and by ERDF/EU, and from Grant No.~GR24022 funded by the Junta de Extremadura (Spain). She is also grateful to the Spanish Ministerio de Ciencia e Innovaci\'on for a fellowship PRE2021-097702 and to Andrés Santos for the fruitful discussions and support during this work.
\end{acknowledgments}

\section*{AUTHOR DECLARATIONS}

\subsection*{Conflict of Interest}

The author has no conflicts to disclose.

\section*{DATA AVAILABILITY}

The data that support the findings of this study are available from the corresponding author upon
reasonable request.

\section*{References}
%

\end{document}